**Single-defect Memristor in MoS$_2$ Atomic-layer**

Saban M. Hus[1*], Ruijing Ge[2], Po-An Chen[3], Meng-Hsueh Chiang[3], Gavin E. Donnelly[4], Wonhee Ko[5], Fumin Huang[4], Liangbo Liang[5], An-Ping Li[5], and Deji Akinwande[1,2*].

[1]Electrical and Computer Engineering, The University of Texas at Austin, Austin, Texas 78758, USA

[2]Texas Materials Institute, The University of Texas at Austin, Austin, Texas 78712, USA

[3]Department of Electrical Engineering, National Cheng Kung University, Tainan, 701, Taiwan

[4]School of Mathematics and Physics, Queen's University Belfast, Belfast BT7 1NN, United Kingdom

[5]Center for Nanophase Materials Sciences, Oak Ridge National Laboratory, Oak Ridge, Tennessee 37831, USA

* Corresponding author: sabanhus@utexas.edu, deji@ece.utexas.edu

**Non-volatile resistive switching, also known as memristor[1] effect in two terminal devices, has emerged as one of the most important components in the ongoing development of high-density information storage, brain-inspired computing, and reconfigurable systems[2-9]. Recently, the unexpected discovery of memristor effect in atomic monolayers of transitional metal dichalcogenide sandwich structures has added a new dimension of interest owing to the prospects of size scaling and the associated benefits[10]. However, the origin of the switching mechanism in atomic sheets remains uncertain. Here, using monolayer MoS$_2$ as a model system, atomistic imaging and spectroscopy reveal that metal substitution into sulfur vacancy results in a non-volatile change in resistance. The experimental observations are corroborated by computational studies of defect structures and electronic states. These remarkable findings provide an atomistic understanding on the non-volatile switching mechanism and open a new direction in precision defect engineering, down to a single defect, for achieving optimum performance metrics including memory density, switching energy, speed, and reliability using atomic nanomaterials.**

The last few years have witnessed significant developments in non-volatile resistance switching (NVRS) devices in a variety of nanoscale material systems such as metal oxides, solid electrolytes, and layered two-dimensional (2D) sheets, where an electric field is employed to switch from high to low resistance states and vice-versa. Among these developments, the demonstration of NVRS in vertical metal-insulator-metal (MIM) devices with single-layer atomic sheets of transitional metal dichalcogenides (TMDs)[10,11] or hexagonal Boron Nitride (hBN)[12] used as the insulating layer (also termed atomristors), is particularly intriguing. While it was long believed that leakage currents would prevent observation of this phenomenon for nanometer thin insulating layers, recent studies on mesoscopic sized atomically-thin crystalline devices demonstrated stable resistive switching. Moreover, NVRS has been observed in a broad portfolio of 2D atomic sheets ($MoS_2$, $MoSe_2$, $WS_2$, $WSe_2$, etc.) with different device geometries and preparation methods indicating that the phenomenon is an intrinsic effect arising from native localized defects and other atomic-scale multi-physics mechanisms. While the atomic scale memory has been demonstrated in several systems before[13,14], it has been restricted to cryogenic temperatures which limit the potential applications. Hence, it is imperative to elucidate the memristor effect in atomically-thin sheets which operate under ambient conditions and can thus enable substantial advances in data storage density, neuromorphic or brain-inspired computing, and radio-frequency reconfigurable communication systems[2,3,11].

Here, by using $MoS_2$ as a model nanomaterial, we report on the atomistic origins of resistive switching in TMD monolayers as being induced by gold adatom substitution of S vacancy defects. With a combination of scanning tunneling microscopy/spectroscopy (STM/STS) and local transport studies, we observe that the S vacancy, which is the predominant defect, does not serve as a low resistance path in its native form. However, metal ions such as the Au migration from the bottom or top electrodes can be substituted into the S vacancy. The resulting metal-occupied defect structure possesses a conducting local density of states (LDOS), and drives the $MoS_2$ atomic sheet to a low resistive state. After removal of the gold atom under a reverse electric field, defects recover their initial vacancy structure and the system returns to a high resistive state.

The pristine experimental samples were prepared by MoS$_2$ exfoliation onto freshly deposited Au surfaces using a previously described gold-assisted mechanical exfoliation technique[15] that yields large-area monolayers (**Fig. 1a**). The gold underlayer served both as a conductive substrate for STM investigations, and as a bottom electrode for in-situ transport studies. Large scale STM images (**Fig. 1b** and **Supplementary Fig. 1**) show continuous MoS$_2$ monolayers that seamlessly cover the terraces and step edges of the underlying gold surface. Raman and photoluminescence (PL) spectroscopy were used to characterize the MoS$_2$ flakes and the acquired spectra are consistent with the characteristics of single-layer MoS$_2$ as displayed in **Fig. 1c and 1d,** respectively[15]. In these spectra, broadening of the PL peak can be ascribed to the variations of the local strain caused by the relative crystal orientation of the underlying gold surface and MoS$_2$ monolayer[15,16]

In order to prepare the samples for atomistic studies, all exfoliated films were annealed at 250°C for several hours in Ultra High Vacuum (UHV) conditions before the initial STM investigations. These mildly annealed samples exhibit a hexagonal lattice of $0.32 \pm 0.02$ nm periodicity of atomically clean MoS$_2$ surface with a very low native defect density (**Fig. 2a**). However, after annealing the samples at 400°C, a large number of defects (in the order of $10^{13}$ cm$^{-2}$) centered around S atom positions is observed (**Fig. 2b**). The dominant defects have been identified as single S vacancies[17,18] ( $V_s$ ), which are known to be the most common in both exfoliated and synthesized MoS$_2$ monolayers[19], due to their low formation energy.

Initial studies began on atomically-resolved regions of MoS$_2$ free of defects. STS measurements on the defect-free regions (**Fig. 2c and Supplementary Fig. 2**) show a bandgap of ~1.4 eV, which is significantly lower than the optical bandgap (~1.8 eV) of monolayer MoS$_2$ and the bandgap values obtained from STS measurements of MoS$_2$ monolayer on HOPG[20]. This reduction of the bandgap has been previously observed and attributed to charge transfer from the gold substrate[21]. While the $V_S$ defects are expected to introduce in-gap states (IGS)[17], STS measurements with stabilization voltages outside the bandgap do not reveal the IGS and do not show a clear bandgap difference between the defect-free regions and the defect regions. However, STS spectra taken with stabilization voltages within the bandgap ($V_{Sample}$=-1.0V) enables the observation of IGS (**Supplementary Fig. 3a**). When a smaller stabilization voltage ($V_{Sample}$= -0.5V) is used, the tunneling principally occurs between the tip and Au substrate, and electronic states of the Au surface can be identified[22,23]. In this case, the STS spectra measured at defect-free regions, and near the defects become identical again (**Supplementary Fig. 3b**). These measurements suggest that $V_s$ defects do not enhance the tunneling current through the $MoS_2$ monolayers or act like conductive filaments in their original configuration.

The traditional STS measurement provides useful information about electron transport through the atomically-thin layers. However, the current density in these measurements is usually limited due to the high tunneling resistance of the vacuum gap between the STM tip and the sample surface. Low current densities and lack of physical contact between the active media and the top electrode may hinder the physical processes that normally occur in MIM memristor devices, such as migration of metal atoms from the electrodes. In order to better replicate the MIM device structure and investigate the NVRS effect at the atomic scale, we employed a gold STM tip as the top electrode for transport measurements. In these experiments, the STM tip was fixed on the surface with a stabilization voltage within the bandgap and approached towards the $MoS_2$ surface for 2-4 Å to form a stable physical contact. Then, DC electrical measurements were performed on this device configuration by sweeping the tip bias.

Transport measurements on defect-free regions reveal a tunneling-like current-voltage (I-V) behavior for low bias voltages (**Fig. 2d**). The I-V curves are asymmetric around zero bias as a result of the two different contact geometries of bottom and top electrodes (flat gold film vs. sharp gold STM tip). Using the same tip and repeating the transport measurements around the defect sites reveal unipolar diode like behavior with a suppressed charge transport for positive tip biases (**Fig 2e**), possibly due to charge accumulation [24] or tip induced modification of band structure[25] around the defect sites. With further tip approaches, the I-V curves become more symmetric for both defect-free regions and at defect positions (**Supplementary Fig. 4**), suggesting a flatter and larger contact between the STM tip and $MoS_2$ surface covering pristine regions around the defects. STM imaging after these transport measurements do not present any visible modification of the MoS$_2$ surface, confirming the generally noninvasive character of low-bias transport measurements with gold STM tips.

While most transport measurements do not alter the MoS$_2$ surface and return I-V curves similar to **Fig. 2d and 2e**, occasionally I-V curves which resemble resistive switching events can be observed at the defect locations (**Fig. 2f**). While it is possible to associate these events to modifications of tip apex, we note that such resistive switching events have never been observed on defect-free MoS$_2$ regions, directly establishing a connection between the S vacancy defects and switching events. STM images before and after such switching events (**Supplementary Fig. 5**) indicate that switching events can initiate the formation of new defects and modifications on the terrace structure of the gold substrate.

In addition, STM investigations reveal a second type of defect, which presents a triangular symmetry (**Fig. 3a**). These defects give a darker contrast than nearby V$_S$ defects and are recognized as S divacancy[26] (V$_{S2}$), the second most energetically-favorable defect in MoS$_2$. Similar to V$_S$ defects, STS measurements on V$_{S2}$ defect sites show a bandgap virtually identical to defect-free regions as presented in **Fig. 3f**. When the transport measurements described above are repeated on V$_{S2}$ defect sites, a switching behavior (a 'SET' event from high to low resistance state) can be triggered at a tip bias ~ 1.8V, where a discontinuity is observed in I-V curves (**Fig. 3b**). After the current reaches the compliance limit (330nA), the bias is decreased, and a hysteresis is observed in I-V curves. STM images taken in the same location after the switching event shows that the dark spot is replaced by a bright protrusion of ~ 0.5 Å height (**Fig. 3c and Supplementary Fig. 6**). The stability and height of the bright spot indicate that the vacancy defect has been occupied by a substituent atom. STS measurements at the bright spot display a non-zero DOS at the Fermi level, showing the metallic characteristics of the defect site (**Fig. 3f**). Furthermore, transport data at low bias voltages indicates that the zero-bias resistance is decreased from 225 MΩ to 50 MΩ. While the measured on/off ratio is relatively small compared to larger micro-meter devices[10], it should be noted that the resistance of the atomic-scale contact between the STM tip and defect site, that is, the contact resistance is likely playing a role in diminishing the intrinsic on/off ratio. Subsequent transport measurements with an opposite tip bias trigger a second resistive switching event from low to high resistance (so-called RESET) at the same defect site where the resistance increases sharply at a tip bias ~ -1.1 V (**Fig. 3d**). Consequent STM imaging in the same location shows that the defect returns to its original vacancy state after the RESET event (**Fig. 3e**).

To elucidate the nature of the defect before and after the switching event, we compared the measured STM dI/dV spectra with the calculated LDOS for an agglomerated $V_{S2}$ vacancy on the opposite sites of $MoS_2$ monolayer and with defect structures where the top S vacancy of $V_{S2}$ complex is substituted with an Au atom. Simulated STM images (**Fig. 4**) qualitatively match the experimental STM images with $V_{S2}$ defects presenting a much higher contrast than $V_S$, and gold substitution defects giving a bright protrusion at the center of the defect site. Importantly, the LDOS calculations (**Fig 4 and Supplementary Fig. 7**) show finite electronic states around the Fermi level after gold substitution in both the $V_S$ and $V_{S2}$ defects, indicating a metallic-like behavior near these gold absorption sites, which were initially insulating. These calculation results corroborate the experimental observation that the gold substitution in the defect site by a positive electric-field introduces conducting electronic states around the Fermi level to reduce the resistance, while the gold removal from the defect site by an opposite field returns the system to the insulating state.

Formation energies of defect structures are critical for determining the stability of these defects and the transitions between them. In $MoS_2$ monolayers, the formation energy for $V_S$ ($V_{S2}$) vacancy varies between 1.3 to 2.9 eV (2.9 to 5.9 eV) for Mo-rich and S-rich limit, respectively (**Supplementary Fig. 8**). While these defect sites are expected to be stable at room temperature and standard STM imaging conditions, due to their relatively low formation energy the S vacancies can still be created via field induced removal of S atoms by the STM tip at high bias voltages and reduced tip-sample distances[27]. Similarly, substitutional gold defects have a binding energy around -3 eV [28,29] per Au atom, which is in the same range of the binding energy of a S atom in a $MoS_2$ monolayer. Enhanced electric fields around the tip apex allow the migration of gold atoms to defect sites or removal of absorbed species from them[30] providing a stable platform for reversible operations. Furthermore, the density of S vacancies can be globally altered by annealing the samples in vacuum or S-rich atmospheres[31]. Therefore, $MoS_2$ monolayers are an ideal platform for a controlled and energetically low-cost formation of vacancy defects to seed switching events.

In summary, we report a one-to-one correlation between NVRS and gold adatom absorption at native vacancy defects of $MoS_2$ monolayers. Due to their inherent layered structure, $MoS_2$ monolayers provide a sharp interface with the metallic top and bottom electrodes. The clean tunneling barriers prevent excessive tunneling currents even in the presence of vacancy defects. However, adatoms from the electrodes can be absorbed in these vacancy sites under a strong electric field. Resulting defect structures provides a low resistive contact between the top and bottom electrodes that changes the resistance state of the memory device. A reversed electric field can remove the adatom from the defect site and return it to its initial state. Such atomic-scale understanding and control of resistive switching of a single defect represents the smallest memristor unit and paves the way for advanced memory applications including extremely high-density memory storage, and brain-inspired neuromorphic computing systems.

## Methods

### Sample Preparation and STM measurements

Very large-scale monolayer $MoS_2$ flakes were mechanically exfoliated on freshly deposited Au films (30 nm Au with 1nm Ti adhesion layer) on 90 nm $SiO_2$ / Si substrates as described previously[15]. The exfoliated $MoS_2$ could be easily identified with an optical camera that guided the STM tip on top of monolayer regions for imaging, spectroscopy, and transport studies. Before the first set of STM measurements, the samples were annealed at T = 250 °C for several hours in UHV conditions ($p < 10^{-10}$ Torr) to remove water and weakly bonded molecules. Initial STM investigations were performed with gold or tungsten STM tips. Samples were later annealed at 400 °C to increase sulfur vacancy density. Afterwards, STM and in situ transport measurements were performed with gold STM tips which were etched using a 50% saturated KCl solution. All STM measurements were made with a variable temperature STM system operated at 100K. For STS measurements a modulation signal of 20 mV at 1Khz was used. For transport measurements, a compliance current of 3.3 nA or 330 nA was used. Before each transport measurement, the gold STM tip was fixed on the surface using a stabilization voltage within the bandgap of $MoS_2$ to ensure a reduced vacuum gap between the tip and $MoS_2$ surface. Then the STM tip was further approached to the surface to provide a stable mechanical and electrical contact. The high mechanical strength of MoS2 prevented any damage to the tip and the sample during the physical contact[25]

### Raman spectroscopy and photoluminescence

Raman spectroscopy and photoluminescence were performed on a Renishaw in-Via system using a 532 nm wavelength source. The typical spot size of the laser beam was around 1.5 μm. While the $MoS_2$ flakes are much larger than the spot size, the size of the randomly oriented gold domains is limited to tens of nm due to thin film growth on amorphous $SiO_2$ surface.

### Computational methods

To simulate STM images of single sulfur vacancy, double sulfur vacancy, and a gold atom adsorbed on top of the vacancy site in monolayer $MoS_2$, we carried out plane-wave DFT calculations by using the VASP package[32]. Projector augmented-wave (PAW) pseudopotentials were used for electron-ion interactions, and the generalized gradient approximation (GGA) with the Perdew-Burke-Ernzerhof (PBE) functional[33] was adopted for exchange-correlation interactions. Monolayer $MoS_2$ was modeled by a periodic slab geometry, where a vacuum separation of 21 Å in the out-of-plane direction (z direction) was used to avoid spurious interactions between periodic images. An 8×8 supercell of monolayer $MoS_2$ (in-plane lattice constant ~25.51 Å) was built to model the sulfur vacancies and gold atom adsorption, where all atoms were relaxed until the residual forces were below 0.01 eV/Å. The plane-wave cutoff energy was set at 280 eV, and a 3×3×1 k-point sampling was used. Based on the optimized structures, we then obtained the partial charge density from VASP for a specific electronic band or at a specific bias energy. Finally, constant-current STM topographic images were computed using the converged partial charge densities within the Tersoff-Hamann approximation[34].

### Data Availability

The data that support the findings of this study are available from the corresponding authors upon reasonable request.

**Acknowledgements**

This work was supported in part by the Presidential Early Career Award for Scientists and Engineers (PECASE) through the Army Research Office (W911NF-16-1-0277), and a National Science Foundation grant (ECCS- 1809017). S.M.H. acknowledges support from a U.S. S&T Cooperation Program. A portion of this research was conducted at the Center for Nanophase Materials Sciences of Oak Ridge National Laboratory, which is a US Department of Energy Office of Science User Facility.

**Author contributions**

S.M.H. conducted STM and transport measurements with the help of W.K. R.G. carried out Raman spectroscopy and photoluminescence measurements. G.E.D. and F.H. prepared large-scale exfoliated monolayer $MoS_2$ samples. P.C. and L.L. performed atomistic simulations. S.M.H., and D.A. initiated the research on atomistic origins of non-volatile resistance switching in single-layer atomic sheets. M.H.C., A.L., and D.A. coordinated and supervised the research. All authors contributed to the article based on the draft written by S.M.H. and D.A.

**Competing interest**

The authors declare no competing financial interest.

**Additional information**

Correspondence and requests for materials should be addressed to D.A.

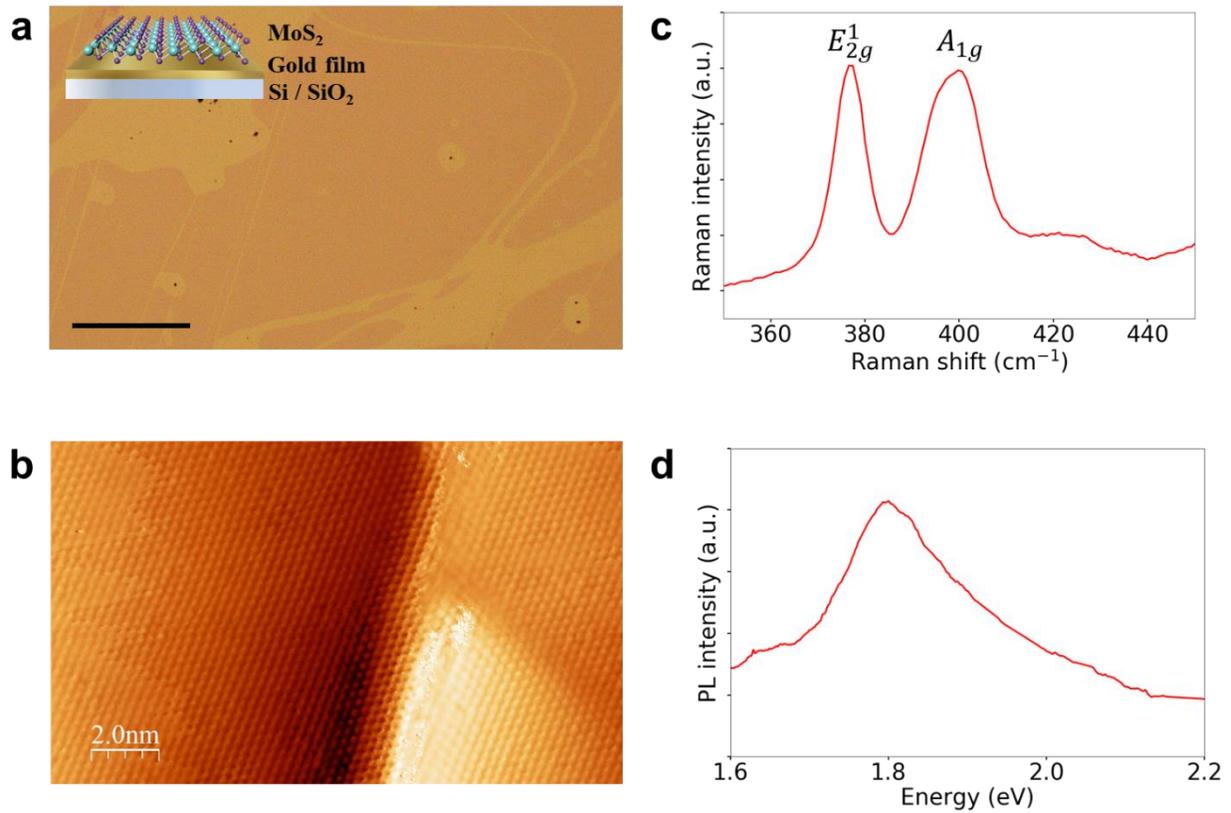

**Fig. 1| Material Characterization**. **a,** Optical image of exfoliated large-area monolayer $MoS_2$ flakes on Au thin film. Scale bar 100 μm. Inset: Schematic of the structure. **b,** STM image showing large, defect-free $MoS_2$ monolayers covering terrace and step edges of the underlying Au film. **c,** Raman spectra of monolayer $MoS_2$ flakes, showing the main in-plane ($E^1_{2g}$) and out-of-plane ($A_{1g}$) vibrational modes. **d,** PL spectra.

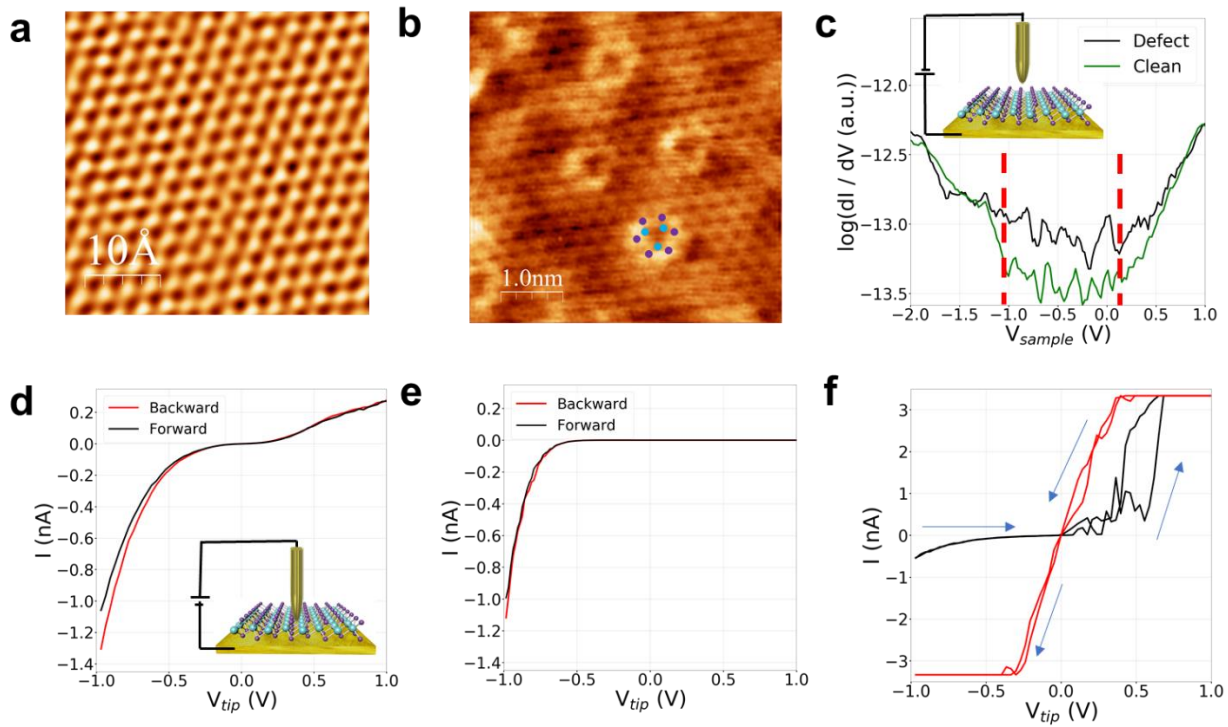

**Fig. 2| Atomistic characterization of MoS$_2$ monolayers**. **a,** Atomically resolved STM images of exfoliated monolayer MoS$_2$ on Au film showing defect-free regions, and **b,** sulfur monovacancy (V$_S$) defects whose density can be increased by vacuum annealing (to the order of $10^{13}$ cm$^{-2}$). Locations of S atoms are shown in purple, Mo atoms in cyan. **c,** dI/dV spectrum taken on a defect-free region is shown in green while the ones taken on and around the defects are shown in black (V$_{Sample}$=-2.5V, I=100pA). Red dashed lines are guides for the eye to mark the bandgap. Inset: Schematic of STM/STS measurements with a vacuum gap for imaging and spectroscopy measurements. **d,** Representative I-V curves for transport measurements on defect-free regions which presents an asymmetric diode like behavior. Inset: Schematic of transport measurements with gold STM tip mimicking atomic-scale metal−insulator−metal device structure. **e,** unipolar diode like I-V curves representing the majority of transport measurements on V$_S$ defect sites. **f,** Resistive switching events observed at two different defect locations during transport measurements.

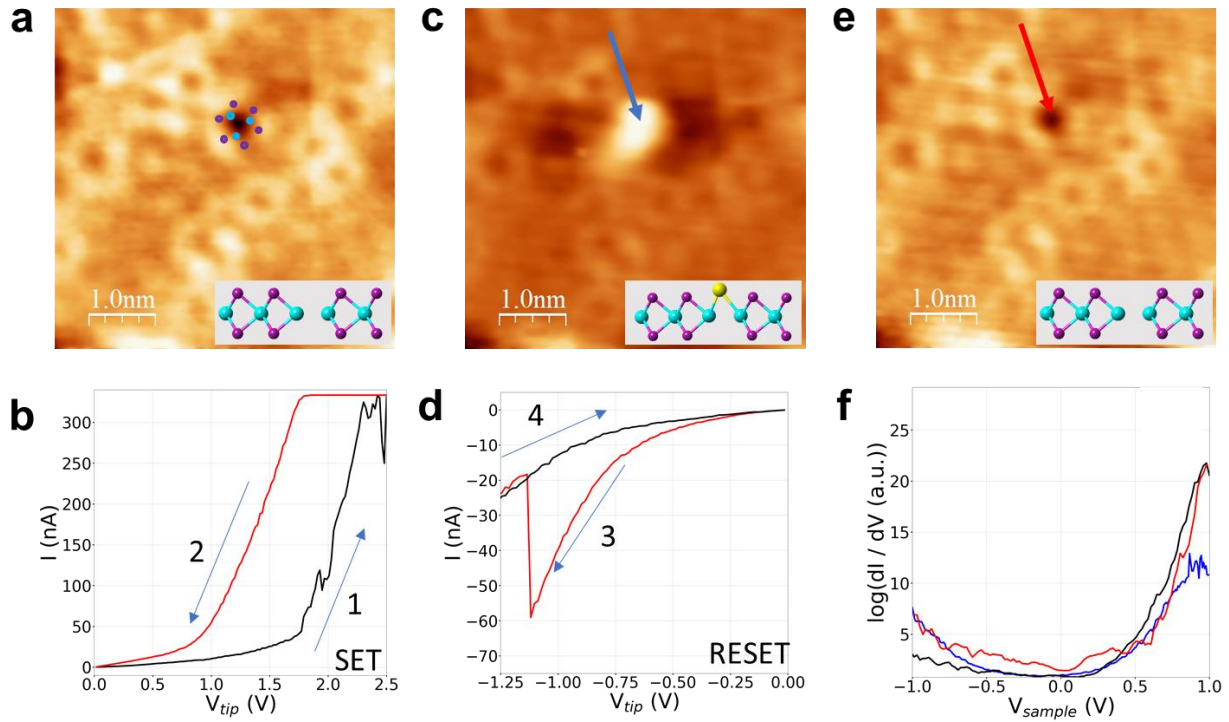

**Fig. 3| Atomistic observation of SET-RESET sequence for V$_{S2}$ defects**. **a,** STM image of the initial V$_{S2}$ point defect. **b,** a SET event induced by voltage bias ~1.8V. **c,** STM image of the same defect after the SET event. Top S site filled by a substitutional gold atom marked with the blue arrow. **d,** RESET event in the defect site. **e,** The defect structure (red arrow) returns to its original configuration (red arrow) after the RESET event. **f,** STS measurements on defect site before (black) and after (red) substitution of gold atom. Compared with the STS measurements taken far away from the defect site (blue curve) (V$_s$=0.7V, I$_t$=100pA).

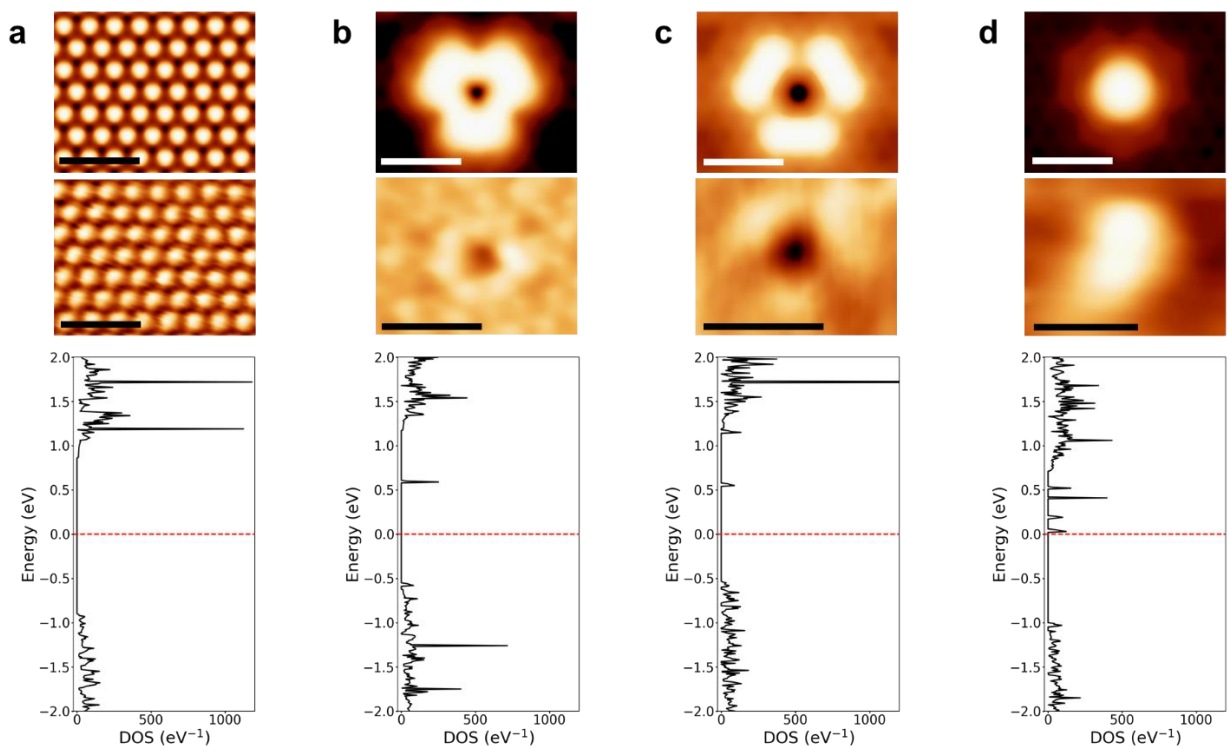

**Fig. 4| Atomistic defect simulations and spectra calculations for monolayer MoS$_2$. a,** Pristine structure. **b,** S monovacancy in the top layer. **c,** S divacancy, and **d,** gold substitution on top S divacancy defect. Top row: Simulated STM images, Middle Row: Experimental images for the same defect structures. Bottom Row: Calculated Local Density of States. The scale bars are 1nm.